# Doping of Self-Standing CNT Fibers: Promising Flexible Air-Cathodes for High Energy Density Structural Zn-air batteries

*Afshin Pendashteh,[a][*] Jesus Palma,[a] Marc Anderson,[a,b] Juan J. Vilatela,[c] and Rebeca Marcilla [a]*

[a] Electrochemical Processes Unit, IMDEA Energy Institute, Parque Tecnológico de Móstoles, Avda. Ramón de la Sagra, 3, 28935 Móstoles, Madrid, Spain.

[b] Department of Civil and Environmental Engineering, University of Wisconsin, Madison, USA.

[c] IMDEA Materials Institute, Tecnogetafe, Calle Eric Kandel, 2, 28906 Getafe, Madrid, Spain.

* afshin.pendashteh@imdea.org

**ABSTRACT:** Finding proper electrocatalysts capable of efficient catalyzing both ORR and OER is of great importance for metal-air batteries. With increasing inclination towards structural and flexible devices, developing a high-performance self-standing air-cathode is highly demanded and challenging, as most of oxygen catalysts are powder and need to be further processed. Here, we construct highly bifunctional air catalyst from macroscopic CNT fibers (CNTf) through direct CVD spinning followed by hydrothermal method. The electrocatalytic properties of the samples were tuned by altering nitrogen-doping and defect densities readily adjusted at different hydrothermal reaction temperatures. The treated CNTfs showed excellent bifunctional activity ($\Delta E=E_{j=10}-E_{1/2}=0.81$ V) and demonstrated exceptional performance as carbon-based self-standing air-cathodes in liquid and solid-state rechargeable Zn-air batteries, with high capacity of 698 mAh·g$^{-1}$ and ultrahigh energy density of 838 Wh·kg$^{-1}$. The rechargeable Zn-air batteries exhibit a low discharge-charge overpotential and excellent stability. This work provides novel simply achieved self-standing air electrodes with exceptional performance for structural Zn-air batteries.

*Keywords:* CNT Fiber; Electrocatalysis; Nitrogen Doping; Self-standing air-cathode; Rechargeable Zn-air Battery; Solid-state Battery.

Tied to the indiscriminate use of fossil fuels, increasing global warming is surely one of the most serious concerns of our century. The Paris climate agreement, signed in December 2015, is proof of the necessity to tackle this issue by reducing emissions. On the other hand, rapid development of portable electronic devices and stationary applications with high energy density requirements urge researchers to develop sustainable compact power sources. Clean energy technologies such as metal-air batteries and fuel cells are promising alternatives for above-mentioned applications.[1] With their open structure and ability to use atmospheric oxygen position metal-air batteries as the most compact and least expensive batteries having extremely high theoretical energy density in comparison with other energy storage technologies.[2] Oxygen reactions are at the center of this technology and are considered as the bottleneck in practical applications. The sluggish kinetics of the oxygen reduction reaction (ORR) under discharge and the oxygen evolution reaction (OER) during the charging of secondary batteries dictates the use of catalysts in order to increase the speed of these reactions.[3] Nobel metals and their alloys (e.g. Pt) are typically used as benchmark catalysts for promoting the ORR while scarce metal oxides (e.g. $RuO_2$ and $IrO_2$) are needed to enhance OER kinetics. Obviously, the high cost and scarcity of such catalysts present difficulties in the commercialization of these renewable energy devices. Therefore, there is a need to explore and develop bifunctional electrocatalysts based on non-precious metals or more preferably metal-free chemistries for these reversible metal-air batteries but also for other electrochemical systems. Many non-precious metal chemistries (e.g. metal oxides or metal sulfides)[4] and their composites with carbon allotropes have been explored as bifunctional catalysts.[5-6] However, metal-free carbons, which show both desirable ORR and OER catalytic activity remains a challenging and important research area of interest. This is due to ease of their preparation, as well as the cost and weight of the final cathode, which directly affects the overall energy density of the device.

Doping of heteroatoms (e.g. N, S, or P) into carbon structures significantly enhances the electrocatalytic activity towards ORR. This enhancement comes from altering the charge density around the carbon atoms adjacent to the heteroatom, which turns them into attractive sites for oxygen adsorption.[7-8] There are also reports showing desirable OER catalytic performance for these heteroatom-doped carbons.[9] Increasing defect density (e.g graphitic edges[10] or defects derived by removal of heteroatoms[11]) and surface functionalization with ketonic groups (C=O)[12] represent other approaches for enhancing catalytic properties of carbon materials.

In practice, electrocatalysts (usually as a powder) must be coated on conductive substrates (e.g. metallic mesh, carbon cloth or paper) and then treated to minimize electrolyte loss. In the case of carbon-based catalysts, utilizing carbon cloth/paper substrates is of more interest due to their lightweight and flexible character. However, these carbon scaffolds do not possess high surface areas or active sites for oxygen reactions, resulting in depressed surface and mass activity.

These factors highlight the importance and attraction of fabricating a 3D self-standing carbon framework on which the active sites are generated which can be used in structural devices.[13-14] Macroscopic carbon nanotube fibers (CNTf) provide such a 3D carbon scaffold, possessing excellent mechanical strength, large surface areas, high conductivity, flexibility, and lightweight character.[15] Recently, we successfully fabricated and employed these fibers and their composites in ionic liquid and all-solid-state supercapacitors.[16-17] These materials displayed excellent electrochemical properties and the ability to be used for structural devices. These excellent features and the fact that CNTfs are easily prepared in large scale suggest that they may be ideal electrodes for metal-air batteries. The caveat to this hypothesis is that one needs to generate sufficient active sites on their surfaces to fulfill this objective. It is known from previous studies that nitrogen doping and increased defect chemistry significantly enhance catalytic activity of carbon nanotubes towards oxygen reactions.[18-19] However, treated CNTs needed to be further processed with binder/additives and be supported on a gas diffusion layer, resulting in complicated air-electrode fabrication and increased dead weight (masses without having catalytic properties). Conversely, this work presents a facile hydrothermal method to engineer defects and doping sites of self-standing CNTfs in a one-step process. The treated fibers play a dual role of providing a 3D self-standing current collector that acts as gas diffusion layer as well as active electrocatalyst sites. This unique characteristic enables these materials to serve as ultralight self-standing air-cathodes with excellent bifunctional catalytic performance towards ORR and OER. This not only simplifies the electrode fabrication process but also removes the necessity of binders, carbon supports or any other additives and makes them ideal air-cathodes for structural devices.

The CNT fibers were fabricated by high-throughput direct CVD spinning, collected as a unidirectional fabric, and finally treated through a facile hydrothermal reaction with urea as the N source. The N-content and defect concentration in carbon structure were readily tuned by altering the treatment condition (e.g. hydrothermal temperature; methods can be found in details in Electronic Supplementary Information, ESI†). The morphological features of the pristine and treated CNTfs were examined by *transmission electron microscopy* (Figure 1a and Figure S1), revealing that the hydrothermal temperature did not significantly modified the morphology of CNTfs. As seen, the fibers are comprised of interconnected CNTs, which are highly crystalline multi-walled with few layers (3-5) and diameter of below 10 nm (inset). Figures S2a and b demonstrate the flexibility and mechanical robustness of the treated CNTfs. The crystallinity of the samples was probed using *Raman spectroscopy* before and after treatment (Figure S3). As seen in the order region, sharp G and D bands appeared for both non-doped and doped samples near 1574 and 1348 cm$^{-1}$, respectively, in accordance with graphitic carbon samples.[20] However, as the treatment temperature is elevated, D band intensity gradually increases, resulting in a slightly higher $I_D/I_G$ ratio. A decrease in crystallinity is consistent with the expected local disruption of the conjugated $sp^2$-hybridised lattice caused by heteroatom inclusion or surface functional groups.

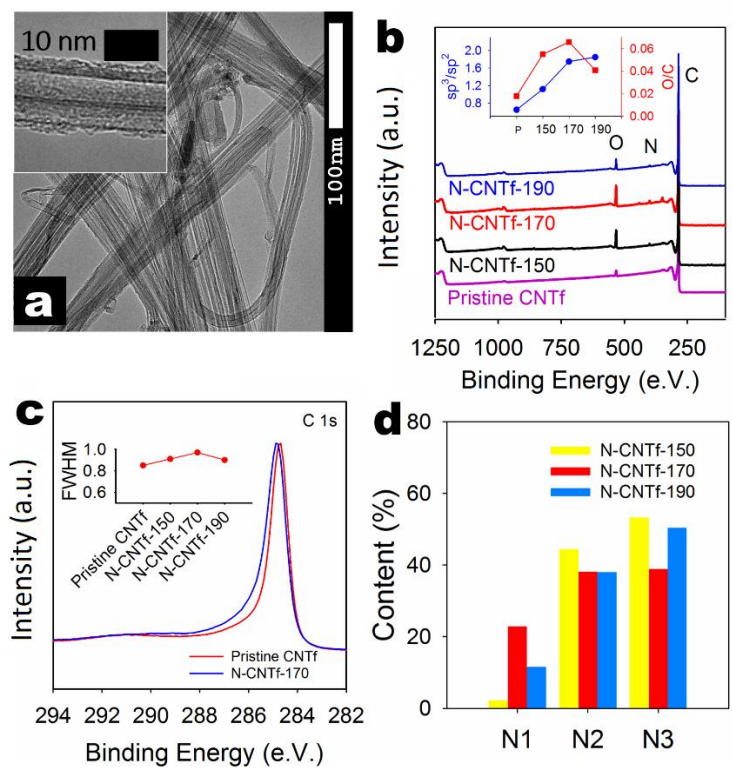

**Figure 1.** (a) TEM images of treated CNT fiber at 170 ℃ (N-CNTf-170). (b) XPS survey spectra of various samples; inset shows the evolution of $sp^3/sp^2$ C and O/C ratio in different samples. (c) High-resolution C 1s spectra of pristine and N-CNTf-170 samples. Inset shows corresponding FWHM evolution for various samples; (d) N 1s core level for various doped samples (N1: pyridinic, N2: pyrrolic, and N3: graphitic).

*X-ray photoelectron spectroscopy* was employed to probe the surface properties and defect/doping nature in treated samples. Figure 1a depicts the survey spectra for pristine CNT fiber and treated samples at various temperatures, revealing the clear emergence of the N 1s peak and significant increased intensity of O 1s after treatment. The increased O content, arisen from the adsorbed functional groups, causes charge redistribution on adjacent carbons. This charge redistribution enhances adsorption of intermediate species in ORR and OER and thus improves the electrochemical properties of the CNTfs.[21] Analyzing C 1s fitting data in inset reveals enhanced $sp^3/sp^2$ and O/C ratios, further suggesting cleavage of the conjugated bonds and appearance of new bonds and functional groups. The C 1s line shifted slightly (0.2 eV) to higher B.E. (Figure 1c) which is in agreement with previous findings on nitrogen doped MWCNT.[22] Moreover, FWHM of C 1s line increases upon hydrothermal treatment, showing a larger contribution in the range of 285-290 eV, which is likely related to C-O and C-N species. More insights were sought through deconvoluting the C 1s signal into five different components (Figure S4): $sp^2$ C (284.6 eV), $sp^3$ C (284.9 eV), C-N/C-O (286.5 eV), C=O (288.8 eV), and the shake-up satellite peak related to $\pi$-$\pi^*$ (290.7 eV), according to our previous assignment.[23] It can be easily realized that the concentration of $sp^3$ C and oxygen functional groups has increased, further suggesting cleavage of the conjugated bonds and appearance of new bonds. This can be further realized in Table S1 where the elemental composition of different samples is listed, exhibiting significant enhanced N and O contents after hydrothermal treatment. Interestingly, this enhancement is more pronounced for the sample treated at 170 ℃ (N-CNTf-170), reaching 1.43%. Furthermore, this sample shows a broader N 1s peak (Figure S5), comprised of three different components: ($N_1$) pyridinic, ($N_2$) pyrrolic, and ($N_3$) graphitic nitrogen. It is clear from peak analysis (Figure 1d) that the N-CNTf-170 possesses the highest concentration of pyridinic nitrogen atoms, which are considered to facilitate oxygen adsorption and hydroperoxide decomposition.[7, 24]

The electrocatalytic activity of non-doped and doped samples was assessed in alkaline solution. Figure S6 displays CV curves of treated and untreated samples. The onset potential in all treated samples was significantly shifted to higher potentials in comparison with pristine CNTf. This clearly demonstrates enhanced catalytic activity of the treated CNTf samples, likely originating from the doped nitrogen atoms, increased defect density and formation of functional groups. Moreover, the N-CNTf-170 sample exhibited the most positive onset potential (0.85 V) and highest current density (1.28 mA·cm$^{-2}$). This is consistent with results obtained from XPS in which N-CNTf-170 sample has a high concentration of pyridinic nitrogen and superficial defects. N-doping increases the density of states near the Fermi level, especially in the case of pyridinic nitrogen, thus facilitating the electron transfer from carbon to O$_2$ σ* antibonding orbitals.[25]

The kinetics of this reaction was further examined by using a *rotating disk electrode* (RDE). Polarization curves of different samples at various rotation rates ranging from 400 to 2300 rpm are shown in Figure S7. By comparing the curves at 1500 rpm in **Figure 2**a one can notice that the current density of the N-CNTf-170 sample experiences a steeper increase in the mixed kinetic-diffusion control region than the other samples including the commercial Pt-based catalyst. This can be attributed to higher N and O content (especially pyridinic nitrogen) in this sample that facilitates the first electron transfer (see ESI†) in the ORR, the rate-determining step.[25]

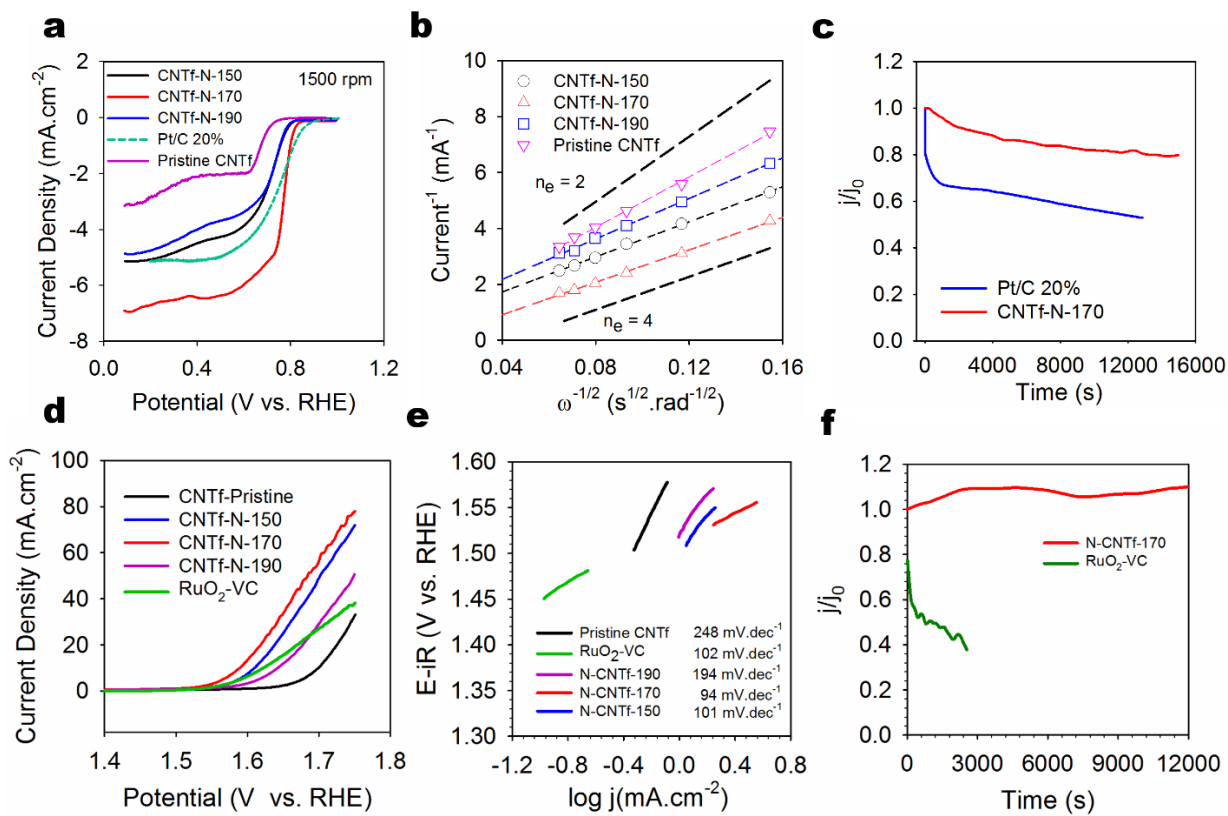

**Figure 2.** Linear sweep voltammograms of pristine and treated CNT fibers and commercial Pt/C 20% at a rotation rate of 1500 rpm with a scan rate of 20 mV·s$^{-1}$ (a). Koutecky-Levich plots for different samples (b), ORR catalytic stability of N-CNTf-170 and commercial Pt/C 20% (c), LSV curves of various samples in Ar-saturated 0.1 M KOH solution, showing their OER catalytic activity (d). Tafel plots of OER LSV curves in panel *d* (e), and OER catalytic durability of N-CNTf-170 in comparison with commercial RuO$_2$-VC (f).

Using the *Koutecky-Levich* theory, the number of transferred electrons ($n_e$) was estimated (Figure 2b). The linearity of the curves suggests a 1$^{st}$-order kinetic reaction towards the dissolved oxygen molecule. Accordingly, $n_e$ was found to be as high as 3.98 for N-CNTf-170 sample (*vs.* 2.87 in pristine CNTf), suggesting that the N-CNTf-170 sample dominantly catalyzes the ORR through a 4*e* pathway just as the commercial Pt-based catalyst. Although there is no consensus on the role of defects, edge plane sites and heteroatom doping on the ORR mechanism, the $n_e$ obtained suggests that the reaction proceeds likely through a two successive 2*e* or a pseudo-4*e* pathway (see

ESI†).[25] The sustained catalytic activity of the N-CNTf-170 sample towards the ORR can be contrasted with that of the commercial catalyst (Figure 2c), demonstrating the outstanding performance of our electrode. The excellent catalytic stability of N-CNTf-170 sample could be attributed to the sturdy structure as confirmed with XRD and TEM after ORR (Figure S8a and c). As seen in XRD, the pattern was remained almost unchanged except that the peak intensities decreased and they are slightly broadened in comparison with the fresh sample, revealing slightly decreased crystallinity after ORR. Just as the ORR is important during discharge, OER is crucial for charging of the metal-air batteries. Accordingly, the electrocatalytic activity with respect to the OER was measured by LSV over a potential range of 1.3 to 1.75 V (Figure 2d). As seen, the N-CNTf-170 sample outperformed the rest of the catalysts, showing an onset potential of 1.54 V. The observed onset is comparable to that of the benchmark $RuO_2$-VC (~1.49 V). In addition, the N-CNTf-170 sample displayed much higher current density in comparison with $RuO_2$-VC. At 1.59 V (an overpotential of only 360 mV) the current raises to 10 mA·cm$^{-2}$, showing the promising OER catalytic properties of the sample (compared to 1.62 V for commercial $RuO_2$-VC catalyst). The superior performance of N-CNTf-170 sample as an efficient OER catalyst can be further gleaned from much a smaller *Tafel* slope of 94 mV·dec$^{-1}$ (e.g. against 102 mV·dec$^{-1}$ for benchmark $RuO_2$) at low over-potentials (Figure 2e). This clearly implies that N-CNTf-170 sample can attain higher current densities at smaller over-potentials. Figure 2f depicts the durability of OER activity for N-CNTf-170 sample and $RuO_2$-based commercial catalyst. It can be clearly seen that treated CNTf did not experience any activity fade while the commercial catalyst lost its activity significantly. This can be correlated to constant composition of the nitrogen-doped CNTf sample in overpotentials needed for OER as suggested by XRD and TEM after OER (Figure S8). Superb electrocatalytic properties of treated CNTf towards OER can be attributed to the insertion of N atoms in the CNTs that induces active sites to facilitate the cleavage of O-H bonds in the water molecule and the formation of O-O bonds in oxygen molecules.[26] Table S2 compares the electrocatalytic performance of treated CNT fibers towards ORR/OER against recently reported free-standing carbon-based catalyst electrodes. The excellent bifunctional catalytic activity of our N-CNT-170 sample can be realized from its small ΔE value (ΔE=$E_{j=10}$-$E_{1/2}$, which is a well-known descriptor for the bifunctional performance of an oxygen electrode). This value comes out at 0.81 V, which is among smallest reported values in literature (Table S2).

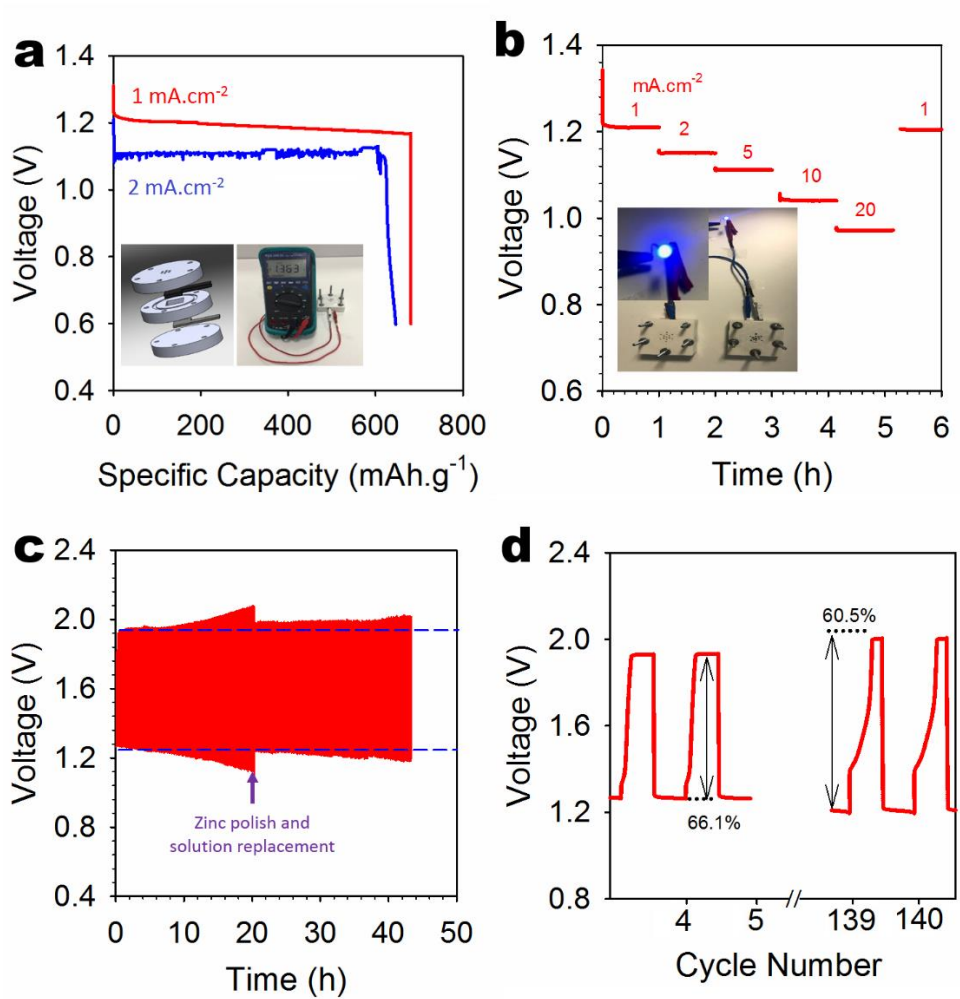

**Figure 3.** (a). Discharge profiles of the Zn-air batteries based on N-CNTf-170 air-cathode at 1 and 2 mA·cm$^{-2}$ up to total Zn exhaustion; inset shows schematic illustration of 3D-printed Zn-air cell and images of a cell with an OCP = 1.36 V. (b) Discharge profiles of the Zn-air batteries at various current densities.

The inset shows powered blue LED by 2 series-connected Zn-air batteries. (c) Long consecutive charge-discharge profile of the Zn-air battery at a current density of 1 mA·cm$^{-2}$ and (d) zoomed areas for initial and final cycles.

The performance of the N-CNTf-170 sample as a freestanding air-cathode was examined by integrating it as cathode with Zn foil as anode, mounted in a 3D printed Zn-air battery cell, as shown in **Figure 3**a. The assembled Zn-air batteries exhibited an open circuit voltage (OCV) of ~1.36 V. As shown in Figure 3a, discharging the cell at a current density of 1 mA·cm$^{-2}$ gives a constant discharge potential of ~1.2 V and a high capacity of 698 mAh·g$^{-1}$ upon total exhaustion of the Zn anode (corresponding to a high energy density of ~838 Wh·kg$_{Zn}$$^{-1}$, around 77.2% of the theoretical energy density of 1086 Wh·kg$^{-1}$). Discharging the cell at 2 mA·cm$^{-2}$ resulted in a discharge voltage of 1.09 V and a capacity of 654 mA·cm$^{-2}$ (~713 Wh·kg$^{-1}$ based on Zn). These remarkable results are the highest capacity and energy values reported for carbon-based catalysts (Table S3), demonstrating outstanding N-CNTf-170 sample performance in ORR/OER catalytic activity. Moreover, discharge profiles of the cell at various current densities ranging from 1 to 20 mA·cm$^{-2}$ have been shown in Figure 3b. As seen, the voltage plateaus are quite stable in different current densities and the device could retain a voltage of 0.97 V at a high current density of 20 mA·cm$^{-2}$. Viability of the assembled Zn-air electrodes was demonstrated by lighting a 5 mm blue LED for more than 1h, as seen in the inset of Figure 3b. In order to probe rechargeability and cycling durability of the N-CNTf-170 sample in Zn-air batteries, long consecutive charge-discharge measurements were performed up to 42 h (Figure 3c). Upon initial cycling, the battery showed only 66% of voltage efficiency (round-trip voltage difference of 0.81 V), which gradually decreases to 56% in cycle 120 (Figure 3d). Recover almost initial voltage difference by repolishing the surface of the Zn foil suggests that the performance fading originated likely from the anode, demonstrating high ORR/OER bifunctional catalytic activity and excellent catalytic durability of N-CNTf-170 air cathode. All of these obtained results suggest that treated CNT fibers can be excellent candidates for lightweight self-standing air-cathodes for flexible solid-state Zn-air batteries. To the best of our knowledge, our self-standing flexible N-CNT fibers are the first freestanding CNT-based catalysts for both ORR and OER, which can be directly employed as an air-cathode.

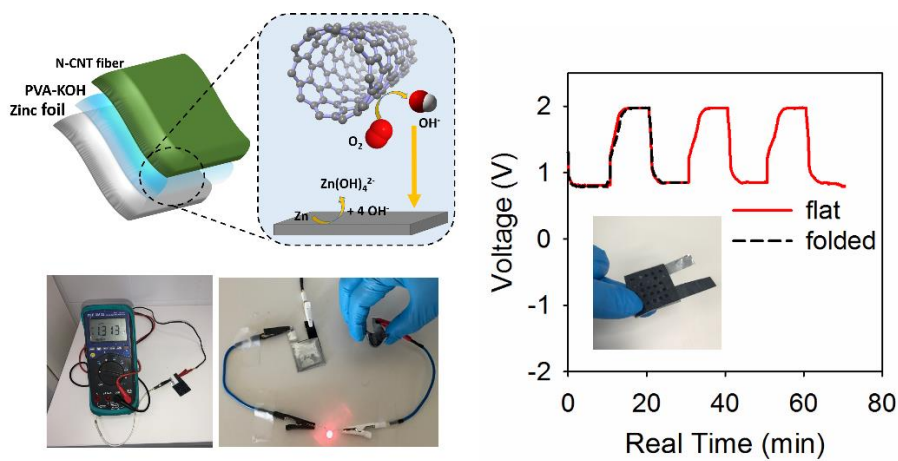

**Figure 4.** Schematic illustration of fabricated rechargeable solid-state Zn-air battery using treated CNT fiber air cathodes and PVA-KOH gel electrolyte (left panel). Right panel shows galvanostatic charge-discharge profiles of the solid-state Zn-air batteries in the flat and bent state.

According to the promising bifunctional catalytic activity of N-CNTf-170 and its performance as an air cathode in liquid-based Zn-air batteries, a rechargeable flexible solid-state Zn-air battery was fabricated (9 cm$^2$) using alkaline poly(vinyl alcohol) gel electrolyte. The solid-state Zn-air battery showed a high OCV of 1.31 V (**Figure 4**), only 50 mV less than the liquid based battery. A schematic illustration of the battery and the reaction mechanism during discharge is shown in Figure 4, whereby the N-CNT-170 catalyzes the ORR through a 4*e* pathway while zinc is oxidized (anode) and zincate species, $Zn(OH)_4^{2-}$ are formed. During charging, the reverse reaction occurs and oxygen is evolved on the treated CNT fibers. Figure 4 displays charge-discharge profiles of the solid-state Zn-air battery, revealing flat voltage plateaus at 1.97 V and 0.87 V during charge and discharge, respectively. It can be also seen that the profile did not change in the bent state. In addition, a red LED could be successfully lit up by two batteries in series while one was bent. This reveals excellent performance of N-CNTf-170 as a flexible air-cathode in the solid device, attributed to outstanding bifunctional catalytic activity of treated CNT fiber electrodes, originated from engineered defect and doping chemistries on the surface of CNT fibers. All of these features along with being lightweight, robust, and easily fabricated on semi-industrial scale make treated CNT fibers excellent air-cathodes for flexible solid Zn-air batteries and wearable/structural devices. This work should inspire further studies on self-standing metal-free air-cathodes for rechargeable Zn-air batteries or broader applications (e.g. Li-$O_2$). However, in addition to optimizing air-cathodes, further extensive studies need to address Zn anode problems and develop flexible and highly ionic conductive solid electrolytes.

In summary, CNT fiber fabrics were easily treated via a hydrothermal route in order to tune their nitrogen doping and defect density. The optimized samples possess high concentrations of pyridinic nitrogen and $sp^3$ carbons, resulting in excellent ORR/OER bifunctional catalytic activity. The treated samples were successfully employed as self-standing air cathodes in rechargeable liquid Zn-air batteries with a high OCP of 1.36 V, a stable discharge voltage of 1.2 V and a high energy density of ~838 Wh·$kg^{-1}$. Moreover, the sample was integrated into a rechargeable high-performance solid Zn-air battery. Accordingly, we believe that the present communication provides a novel concept of doping and defect engineering of carbon fabrics, which we feel can be employed for further development of advanced self-standing air cathodes for structural metal-air batteries.

## ASSOCIATED CONTENT

**Supporting Information.**

Experimental section, photographs, SEM images, RAMAN spectra, high-resolution XPS core-levels of C 1s and N 1s, cyclic voltammograms, LSV curves at various rotation rates for different samples. Tables included: (S1) Surface elemental composition of various samples obtained from XPS, (S2) Comparison of the bifunctional catalytic properties of N-CNTf-170 with recently reported carbon-based flexible electrodes, and (S3) Comparison of the Zn-air battery performance using N-CNTf-170 as air-cathode with recently reported ZABs using carbon-based flexible electrodes. This material is available free of charge via the Internet at http://pubs.acs.org.

AUTHOR INFORMATION

**Corresponding Author:**

afshin.pendashteh@imdea.org  ORCID: 0000-0002-9884-3721

**Author contributions:**

The manuscript was written through contributions of all authors. All authors have given approval to the final version of the manuscript.

**Notes:**

The authors declare no competing financial interest.

**ACKNOWLEDGEMENT**

Authors gratefully acknowledge financial support from MINECO (former MICINN) through the MAT2015-64167-C2-1-R project, European Union structural funds and the Comunidad de Madrid MAD2D-CM Program (S2013/MIT-3007). JJV is grateful for generous financial support provided by the European Union Seventh Framework Program under grant agreements 678565 (ERC-STEM) and by MINECO (RyC-2014-15115). In addition, authors would like to thank *Dr. Belén Alemán* for insightful discussion and advice on XPS interpretation.**REFERENCES:**

1. Lin, L.; Zhi-wen, C.; Xin-Bo, Z., Recent Progress on the Development of Metal-Air Batteries. *Advanced Sustainable Systems* **2017,** *1*, 1700036.
2. Wei, L.; Karahan, H. E.; Zhai, S.; Liu, H.; Chen, X.; Zhou, Z.; Lei, Y.; Liu, Z.; Chen, Y., Amorphous Bimetallic Oxide–Graphene Hybrids as Bifunctional Oxygen Electrocatalysts for Rechargeable Zn–Air Batteries. *Advanced Materials* **2017,** *29*, 1701410-n/a.
3. Liang, Y.; Li, Y.; Wang, H.; Zhou, J.; Wang, J.; Regier, T.; Dai, H., Co3o4 Nanocrystals on Graphene as a Synergistic Catalyst for Oxygen Reduction Reaction. *Nat. Mater.* **2011,** *10*, 780-786.
4. Pendashteh, A.; Palma, J.; Anderson, M.; Marcilla, R., Nicomno4 Nanoparticles on N-Doped Graphene: Highly Efficient Bifunctional Electrocatalyst for Oxygen Reduction/Evolution Reactions. *Appl. Catal., B* **2017,** *201*, 241-252.
5. Meng, F.; Zhong, H.; Yan, J.; Zhang, X., Iron-Chelated Hydrogel-Derived Bifunctional Oxygen Electrocatalyst for High-Performance Rechargeable Zn–Air Batteries. *Nano Research* **2017,** *10*, 4436-4447.
6. Hai-Xia, Z.; Jun, W.; Qi, Z.; Fanlu, M.; Di, B.; Tong, L.; Xiao-Yang, Y.; Zhi-Wen, C.; Jun-Min, Y.; Xin-Bo, Z., In Situ Coupling Fem (M = Ni, Co) with Nitrogen-Doped Porous Carbon toward Highly Efficient Trifunctional Electrocatalyst for Overall Water Splitting and Rechargeable Zn–Air Battery. *Advanced Sustainable Systems* **2017,** *1*, 1700020.
7. Wiggins-Camacho, J. D.; Stevenson, K. J., Mechanistic Discussion of the Oxygen Reduction Reaction at Nitrogen-Doped Carbon Nanotubes. *J. Phys. Chem. C* **2011,** *115*, 20002-20010.
8. Wiggins-Camacho, J. D.; Stevenson, K. J., Effect of Nitrogen Concentration on Capacitance, Density of States, Electronic Conductivity, and Morphology of N-Doped Carbon Nanotube Electrodes. *J. Phys. Chem. C* **2009,** *113*, 19082-19090.
9. Zhao, Y.; Nakamura, R.; Kamiya, K.; Nakanishi, S.; Hashimoto, K., Nitrogen-Doped Carbon Nanomaterials as Non-Metal Electrocatalysts for Water Oxidation. *Nature Communications* **2013,** *4*, 2390.
10. Stamatin, S. N.; Hussainova, I.; Ivanov, R.; Colavita, P. E., Quantifying Graphitic Edge Exposure in Graphene-Based Materials and Its Role in Oxygen Reduction Reactions. *ACS Catal.* **2016,** *6*, 5215-5221.
11. Jia, Y.; Zhang, L.; Du, A.; Gao, G.; Chen, J.; Yan, X.; Brown, C. L.; Yao, X., Defect Graphene as a Trifunctional Catalyst for Electrochemical Reactions. *Advanced Materials* **2016,** *28*, 9532-9538.
12. Lu, X.; Yim, W.-L.; Suryanto, B. H. R.; Zhao, C., Electrocatalytic Oxygen Evolution at Surface-Oxidized Multiwall Carbon Nanotubes. *J. Am. Chem. Soc.* **2015,** *137*, 2901-2907.
16